\documentclass{optica-article}
\journal{opticajournal} % for journals or Optica Open
\articletype{Research Article}
\usepackage{changepage}
\usepackage{multirow}
\usepackage{booktabs}
\usepackage{tabularx}
\usepackage{bm}
\usepackage{color}
\usepackage{cite}
\usepackage{amsmath}
\usepackage[figuresright]{rotating}
\begin{document}
\title{Block definition design for stretchable metamaterials: enabling configurable sensitivity to deformation}
\author{Sihong Chen,\authormark{1,2,3} Taisong Pan,\authormark{1,*} Zhengcheng Mou,\authormark{4} Mingde Du,\authormark{3} Tianxiang Wang,\authormark{2} Bing-Zhong Wang, \authormark{2} and Yuan Lin\authormark{1,5}}

\address{\authormark{1}School of Materials and Energy, University of Electronic Science and Technology of China, Chengdu 610054, China\\
\authormark{2}School of Physics, University of Electronic Science and Technology of China, Chengdu 610054, China\\
\authormark{3}Huawei Technologies, Chengdu 610041, China\\
\authormark{4}School of Statistics, Southwestern University of Finance and Economics, Chengdu 611130, China\\
\authormark{5}Medico-Engineering Cooperation on Applied Medicine Research Center, University of Electronic Science and Technology of China, Chengdu 610054, China}

\email{\authormark{*}tspan@uestc.edu.cn} %% email address is required; see note below about the corresponding author designation

% use {asbstract*} to suppress the copyright line. Copyright information will be added in production
\begin{abstract*} 
The sensitivity to deformation plays a key role in determining the applicability of stretchable metamaterials (MMs) to be used for conformal integration or mechanical reconfiguration. Typically, different unit designs are required to achieve the desired sensitivity, but this article proposes a block definition design for stretchable MMs that enables regulation of the MMs' response to deformation by defining various block arrangements with the same precursor structure. The article demonstrates a stretchable MM that employs the block definition design to show the mechanical reconfigurability of resonant frequency. Different block definitions result in modulation ranges of resonant frequency ranging from 39\% to 85\% when applying a 20\% tensile strain. Additionally, the proposed design is also used to realize another MM with contradictory sensitivity to the deformation and electromagnetically induced transparency (EIT) MMs with configurable transmission bandwidth to the deformation, indicating its potential for broader applications.
\end{abstract*}

%Stretchable metamaterials, block definition, conformal integration, mechanical reconfiguration, configurable sensitivity
%%%%%%%%%%%%%%%%%%%%%%%%%%  body  %%%%%%%%%%%%%%%%%%%%%%%%%%
\section{Introduction}
Metamaterials (MMs) have demonstrated diverse applications in controlling electromagnetic (EM) wave, including perfect absorber, EM cloaking and hologram imaging\cite{landy2008perfect,liu2010infrared,tittl2015switchable,durmaz2019polarization,wang2019large,schurig2006metamaterial,zhang2019phase,qian2020deep, wan2020holographic,cui2020information,guo2019reconfigurable}. Recent advances in stretchable MMs have made them even more versatile, allowing them to bend, stretch, and twist. With this ability of mechanical deformation, two primary applications of stretchable MMs have emerged: conformal integration\cite{jiang2011conformal,cong2014highly,zhou2018stretchable,hashemi2019flexible,lee2012reversibly,liang2015anomalous,yang2016flexible,ee2016tunable,Gurrala2017Fully} and mechanical reconfiguration \cite{malek2017strain,nauroze2018continuous,pryce2010highly,lee2019single,choi2016electroactive}. When utilizing stretchable MMs for conformal integration, they conform to the target surface by deforming accordingly. By attaching them to the surface, their stretchability ensures their shape is maintained, even on complex surfaces. Rather than adapting to surfaces through deformation, mechanical reconfiguration intentionally applies specific strain to stretchable MMs to manipulate their performance via desired deformation. Although there is still a gap of switching time between mechanical and electrical reconfiguration, reconfiguring MMs through mechanical deformation offers advantages such as not requiring external power or additional components.

However, the two application scenarios for stretchable MMs have conflicting requirements for their response to deformation. For conformal integration, when attached to a complex surface, stable performance before and after integration is expected, but for mechanical reconfiguration, performance should be highly dependent on deformation. This means that conventional design approaches require separate designs for each scenario, leading to design complexity. Thus, the use of a single MM structure for both scenarios could simplify the application of stretchable MMs. Additionally, fabricating stretchable MMs for gigahertz (GHz) and megahertz (MHz) regimes can be challenging due to the larger feature sizes, which often range from millimeters to centimeters, resulting in a larger overall size compared to those used in terahertz (THz) and optical applications. The conventional fabrication method for stretchable MMs involves pre-patterning discrete unit cells on a rigid substrate, which are then transfer-printed onto an elastomer substrate \cite{lee2012reversibly,kamali2016decoupling,yu2013stretchable,ni2015ultrathin,geiger2020flexible}. However, this method faces challenges such as misalignment and low yield when preparing large-scale MMs due to the non-uniformity of the applied force and adhesion conditions during the transfer process.

Thus, this study proposes a block definition design strategy to address difficulties encountered in the realization of stretchable MMs for radio-frequency applications. Instead of using the conventional "pattern-transfer" fabrication method, the block definition design utilizes the "transfer-define" method. In this method, a unified precursor structure is prepared and transferred to the elastomer substrate in its entirety. The stretchable MM is then obtained by partitioning the precursor into several blocks using ultraviolet laser micromachining. The MM's properties can be tuned using different block definitions, and the overall transfer of the precursor structure simplifies the fabrication of MMs with large-scale unit cells. To demonstrate the feasibility of this design strategy, a stretchable MM was fabricated and evaluated. The fabricated MM, which featured tightly-coupled square-loop elements, exhibited a strong frequency selective effect in the frequency range of 2 - 18 GHz, and its sensitivity to tensile strain could be adjusted using different block arrangements within the 0\% to 20\% strain range. Detailed electric field analyses were conducted to explain the MM configuration mechanism with block definition design. Finally, to illustrate the broader applicability of the block definition design, another MM design with opposite frequency response sensitivity and an electromagnetically induced transparency (EIT) MM with a configurable bandwidth dependence on deformation were presented.

\section{Design concept and fabrication process}
The basic concept of block definition design for stretchable MM is illustrated in Figure 1. To define a stretchable MM, the precursor structure is partitioned into periodically arranged blocks on a soft substrate. This partitioning enables the MM to deform under tensile strain while also determining its fundamental properties and sensitivity to deformation. By adjusting the partitioning strategy, it is possible to reconfigure the MMs' response to deformation based on the same precursor.  
\begin{figure*}[!h]
	\centering
	\includegraphics[width=4.2in]{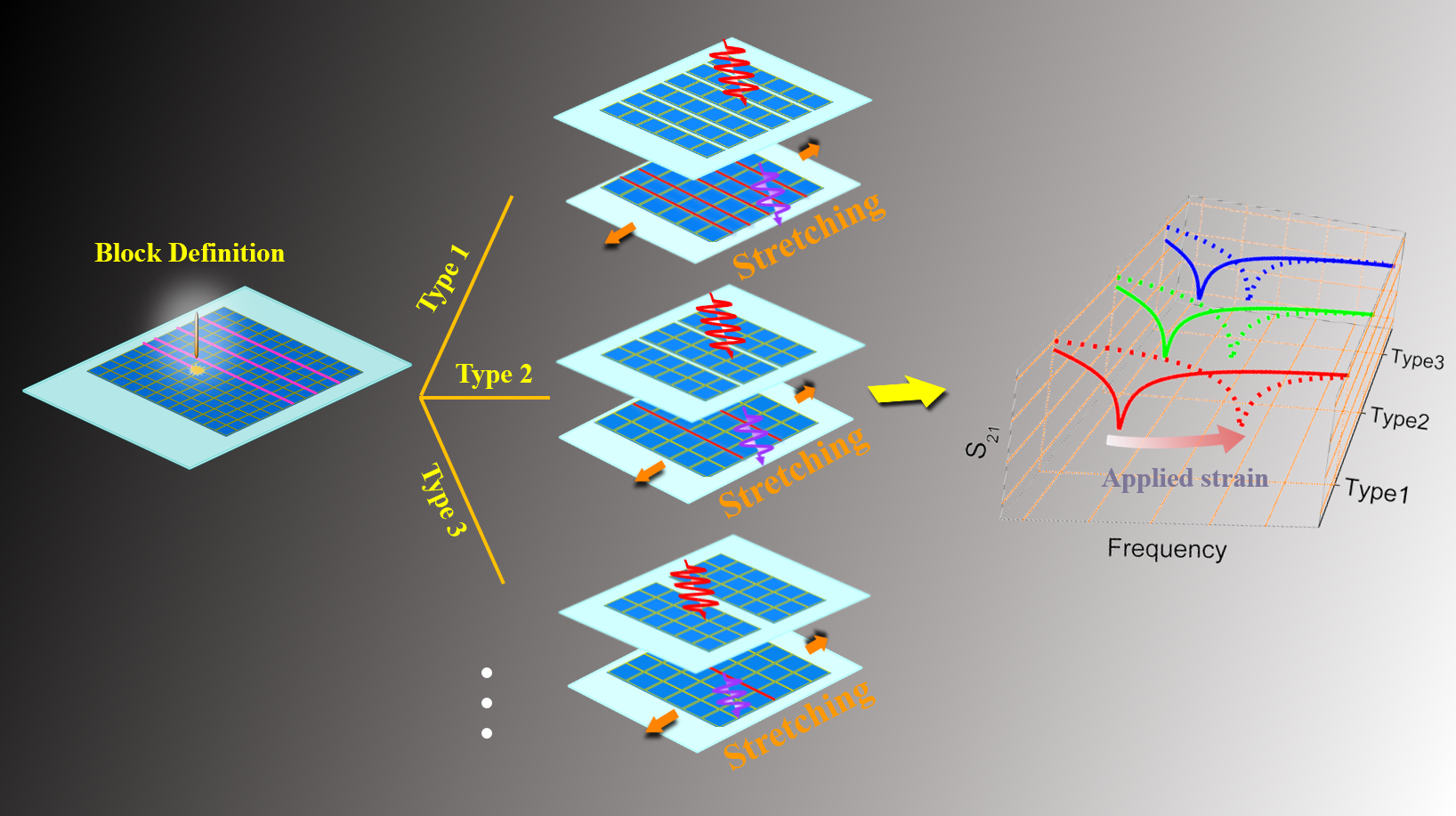}
	\caption{The diagram demonstrates the use of block definition design in creating a stretchable MM. This approach involves dividing the precursor structure into distinct blocks, enabling MMs to stretch and respond differently to identical strain loads using the same precursor.}
	\label{figure1}
\end{figure*}

To demonstrate its realizability, a square-loop structure is selected as the building block of the precursor due to its extensive research as the unit cell for MMs \cite{yilmaz2009design,parker1991gentleman,monacelli2005infrared}. 
The square-loop structure has dimensions of $a$ $\times$ $a$ and a periodicity of $P_x$ $\times$ $P_y$, as shown in Figure 2(a). The precursor consists of an array of 40 $\times$ 40 square-loop unit cells with an initial distance of $g_x$ and $g_y$, where the line width of each unit cell is $w$. 
Detailed parameters can be found in Table \ref{table_fss1}. Figure 2(b) illustrates the fabrication process, which uses the "transfer-define" method to create the precursor with a flexible copper clad laminate (FCCL), which consists of a 18-$\mu$$m$-thick copper layer and a 110-$\mu$$m$-thick polyimide (PI) layer. As the metal patterns on the FCCL can be completed by the conventional flexible printed circuit technique, the MM precursor with a total size of $200$ $\times$ $200$ $mm^2$ can be easily fabricated with high yield. 

The precursor is then transferred in its entirety to an adhesive polyurethane (PU) substrate. An in-situ define technique is employed using an ultraviolet laser to complete the definition of block assembly and form the stretchable MM. As the ultraviolet laser has no significant local heating effect, it can cut the precursor structure into several blocks without damaging the elastomer substrate by adjusting the cutting times. Figure 2(c) demonstrates the dependence of cutting depth on cutting times with a laser power of 1.5 $mW$, indicating that insufficient cutting times (3 times) lead to the defined blocks being unable to be separated, whereas excessive cutting (7 times) causes cracks to appear in the PU substrate. Given that the PI layer of the FCCL is about 0.11-$mm$-thick, the optimized cutting times of 5 are determined. Figure 2(d) displays the as-fabricated MM with different block definitions, including type-1, type-2, and type-3. These types are defined based on the same precursor of MM and consist of 1, 2, and 3 columns of square-loop elements, respectively, for each block of the MM.
\begin{figure*}[!h]
	\centering
	\includegraphics[width=4.5in,height=3.4in]{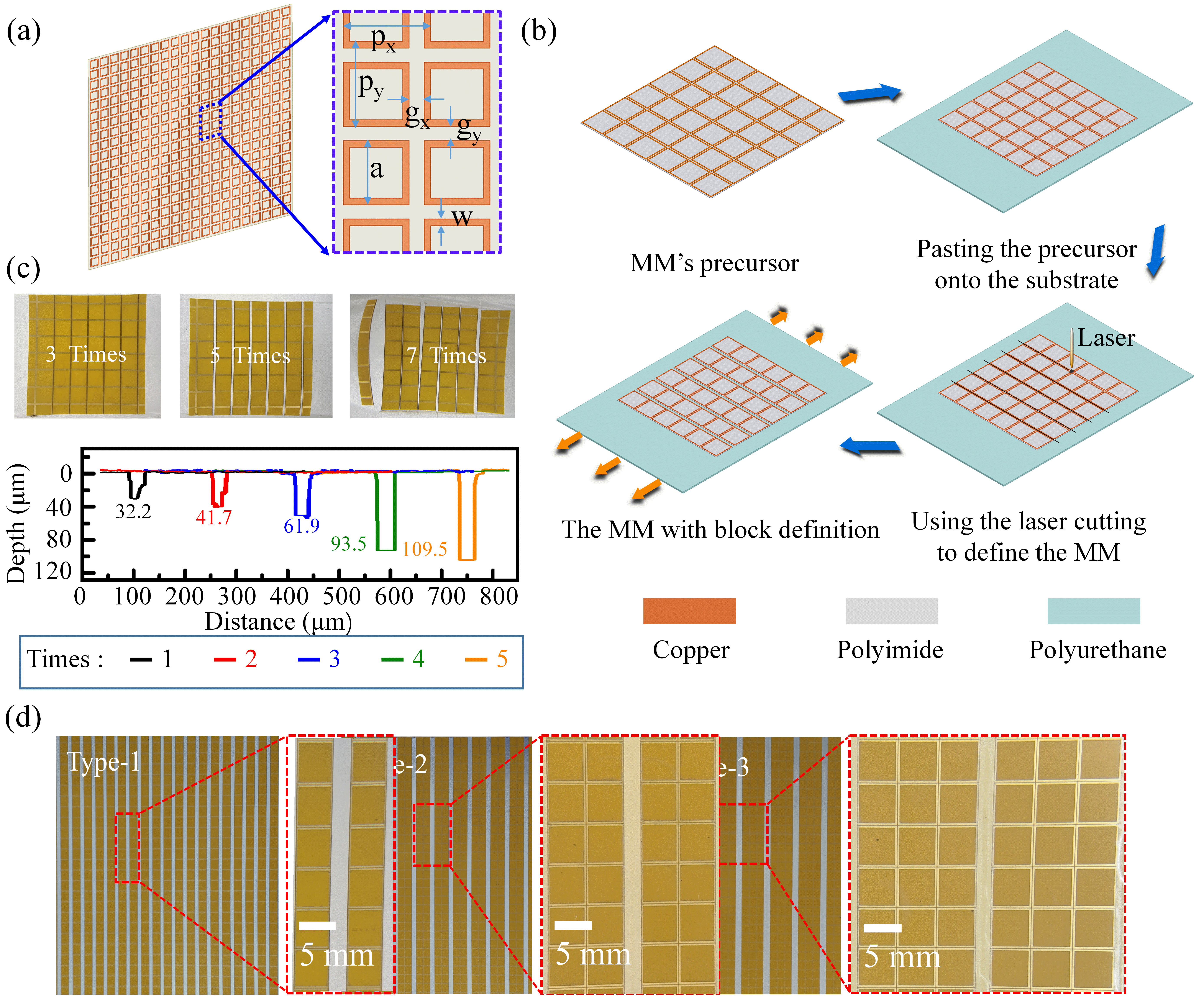}
	\caption{(a) Geometric illustration of the precursor formed by square-loop elements. (b) Fabrication process with the “transfer-define” method. (c) Dependence of the cutting depth on the cutting times with the laser power of $1.5\ mW$ and corresponding photographs of the precursors defined by different cutting times. (d) Photographs of the stretched MMs with type-1, type-2 and type-3 definitions.}
	\label{figure2}
\end{figure*}

\begin{table}[!h]
	%% increase table row spacing, adjust to taste
	\renewcommand{\arraystretch}{1.1}
	% if using array.sty, it might be a good idea to tweak the value of
	\caption{Parameters of the MM consisted of square loop}
	\label{table_fss1}
	\centering
	\begin{tabular}{|c|c|c|c|c|c|}
		\hline
		$g_x$  & $g_y$  & $a$ & $w$ & $P_x$  & $P_y$ \\
		\hline
		0.1	$mm$&  0.1 $mm$&4.9 $mm$ & 0.2 $mm$ & 5.0 $mm$  & 5.0 $mm$ \\
		\hline
	\end{tabular}
\end{table}

The electromagnetic response of the MMs was studied using the finite element method. To emulate an infinite structure and save computational resources, master/slave boundary conditions and Floquet port excitation were applied in the simulation. The transmission responses of the fabricated MMs were measured using a Vector Network Analyzer (Rohde \& Schwarz ZNB20). The measurement setup is depicted in Figure 3, where two ridge antennas operating at 2-18 GHz were used and placed in the line of sight with each other. The MM sample was placed in between the two antennas, with a distance far enough to ensure that the sample was in the far-field region and excited by a plane wave. Calibration was performed before conducting the measurements to reduce test error. The actual transmission response was obtained by subtracting the response without the MM from the one with the MM. The required tensile strain is manually applied to the MMs by controlling the stretching distance using a homemade stretching holder at a loading speed of approximately 1 cm/s. Other actuation methods such as motor drive and shape memory polymers could be used to accelerate the switching time of the proposed MM.
\begin{figure}[!h]
	\centering
	\includegraphics[width=3in]{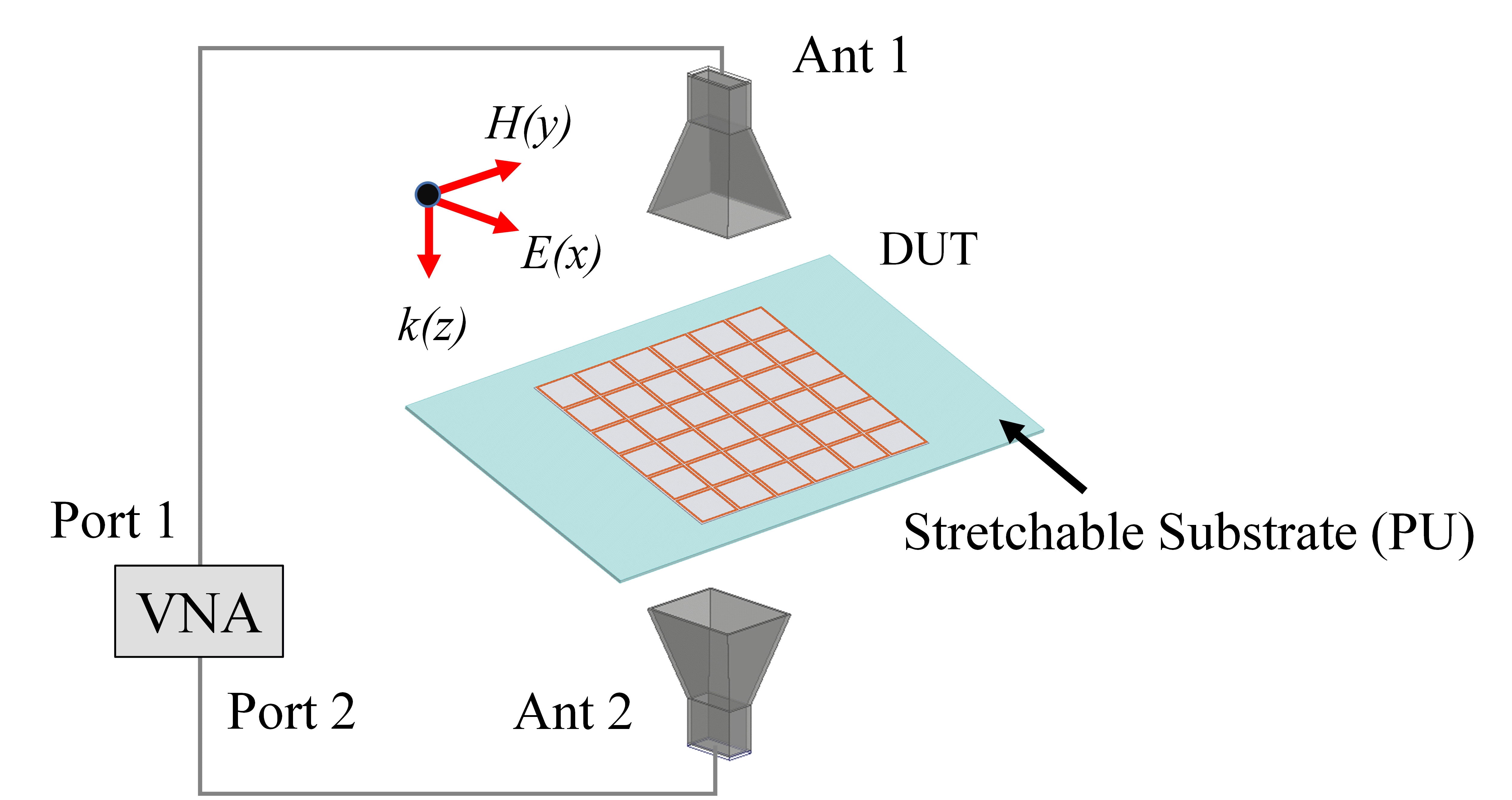}
	\caption{Geometrical illustration of experimental setup for the stretchable MMs.}
	\label{fig3}
\end{figure}

\section{Simulation and experiment results}
\begin{figure*}[!h]
	\centering
	\includegraphics[width=4.5in,height=4.2in]{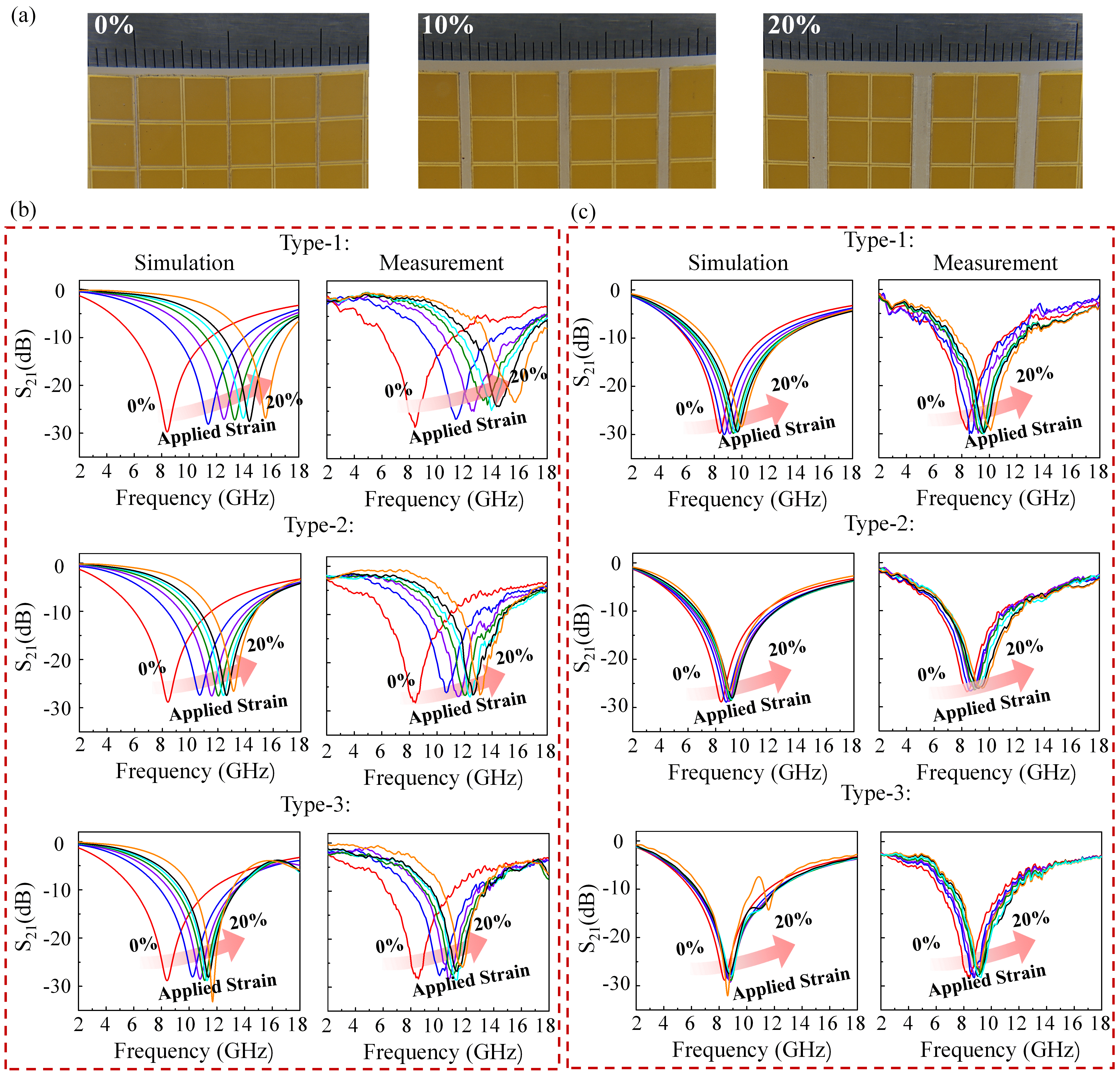}
	\caption{(a) Photographs of the type-2 MM with different applied strains. The simulated and measured transmission spectra of TE polarization (b) and TM polarization (c) for type-1, type-2 and type-3 MMs with different applied strains.}
	\label{figure3}
\end{figure*}

To examine how block definition affects transmission response to deformation, we first investigate the dependence of the $S_{21}$ parameter on applied tensile strain when an electromagnetic wave is incident normally. As shown in Figure 4(a), applying tensile strain increases the distance between defined blocks, while the structure of each block remains the same as the MM deforms. Figure 4(b) displays transmission spectra for MMs with type-1, type-2, and type-3 definitions, obtained from simulations and experiments with TE polarization. The results demonstrate high consistency between the simulations and experiments. All three types of MMs exhibit typical frequency selective effects over the frequency range of 2 GHz to 18 GHz, and the frequency selective effect persists when the MM is stretched with tensile strain. To avoid plastic deformation of the PU substrate \cite{wang2019strain, dal2021performance}, the maximum strain is set to 20\%. When tensile strain is applied, the resonant frequencies of all three types of MMs shift to higher frequencies from 8.4 GHz. However, as the block design changes from type-1 to type-3, the amplitude of the frequency shift gradually decreases, indicating that the response of the resonant frequency to deformation can be effectively tuned by defining the columns of square-loop elements in each block. Meanwhile, there is no significant difference in bandwidth. Additionally, transmission spectra for MMs with TM polarization are presented in Figure 4(c). In comparison with results for TE polarization, both the resonant frequency and bandwidth of the MM are less sensitive to deformation. This difference in polarization sensitivity can be attributed to the asymmetrical deformation of the MMs under applied uniaxial strain.

%\end{adjustwidth}
To further illustrate the significant impact of block definition on the MMs' response to deformation. Figures 5(a) and 5(b) plot the dependence of resonant frequency shift and bandwidth on deformation for the three types of MMs. Simulation and experimental results demonstrate that the sensitivity of resonant frequency shift to deformation decreases markedly as the number of columns in each block increases, from type-1 to type-3. In the case where a single block of the MM consists of only one column of square-loop structure (type-1), the resonant frequency shift is 85\% (from 8.4 GHz to 15.57 GHz) for a 20\% applied strain. By contrast, defining the block as two or three columns of square-loop structure (type-2 or type-3, respectively) results in gradually smaller resonant frequency shifts with the same 20\% tensile strain, namely 44.5\% (from 8.4 GHz to 12.14 GHz) and 39\% (from 8.4 GHz to 11.68 GHz), respectively. However, the block definition has little effect on the dependence of MM bandwidth on deformation. When the MMs are stretched by 20\% tensile strain, the bandwidths of the MMs with all three block definitions decrease gradually by approximately 2 GHz. 

The close relationship between block definition and frequency selective effect simplifies the design process for conformal attachment and mechanical modulation. The proposed MMs can meet the demands of both scenarios with a single precursor structure, eliminating the need for separate designs. For instance, when considering conformal attachment using the proposed MMs, type-3 definition (three columns of square-loop structure in each block) may be employed to cut the precursor structure and achieve MMs with minimal resonant frequency shifts after deformation. On the other hand, the type-1 definition (one column of square-loop structure in each block), which exhibits significant dependence of resonant frequency on deformation, is highly suitable for realizing mechanical reconfiguration of MMs' transmission properties

\begin{figure}[!h]
	\centering
	\includegraphics[width=4in,height=4in]{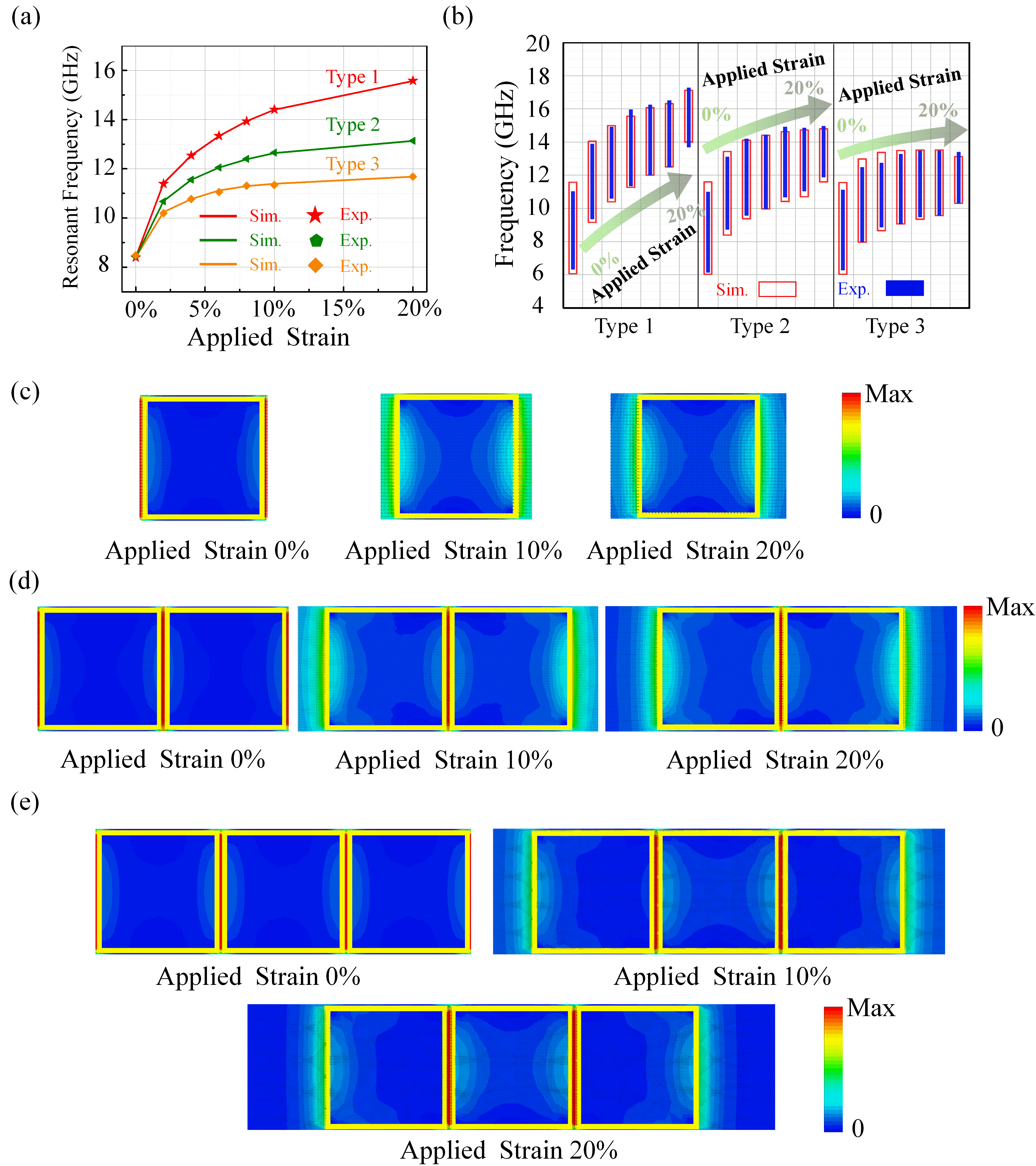}
	\caption{(a) Dependence of the resonant frequency on the applied strain when the proposed three types of MMs are illuminated by TE-polarized wave. (b) Comparison of simulated and measured bandwidth of the MMs with different applied strain. (c-e) Electric field distributions of the proposed MMs under TE polarization with applied strain of 0\%, 10\%, and 20\%, respectively.}
	\label{figure4}
\end{figure}

To explain how the MMs respond to deformation based on block definition design, we analyzed the electric field in detail. Figures 5(c-e) illustrate the electric field distributions for type-1, type-2, and type-3 definitions under TE polarization, respectively, with applied strain varying. In Figure 5(d), when the MM with type-2 definition is undeformed, concentrations of the electric field can be observed around the interface between the two columns of square-loop elements and the cutting slits (leftmost and rightmost sides of the block). These current concentrations indicate strong coupling between the blocks and the two columns of square-loop elements in each block. Upon the application of tensile strain, the electric field magnitude around the cutting slits gradually decreases, while that in the interface of the columns remains strong. Similar trends are observed in MMs with type-1 and type-3 definitions in Figure 5(c) and Figure 5(e). These simulation results of electric field distribution with several deformation states provide a preliminary understanding of the mechanism of regulating the transmission properties of the stretchable MM with block definition design. The elongation of the MM and the increase of the distance between the separated blocks due to applied tensile strain result in a decrease in the coupling strength between the blocks, which is sensitive to the applied strain. However, the coupling between the unit cells in each block remains stable during the deformation of the MM. By partitioning the precursor structure with different block definitions, the ratio of "movable" and "fixed" unit cells can be adjusted, and the proportion of the unit cells under stable coupling conditions can be tuned. This results in different responses of the MMs to deformation.

\section{Further applications of block definition design} 
Two additional stretchable MMs with block definition design are proposed to showcase the wide-ranging applications of this design approach. One of these is a stretchable MM with configurable resonant frequency shift, while the other is an MM with configurable EIT effect. 
\begin{figure}[!h]
	\centering
	\includegraphics[width=4.5in]{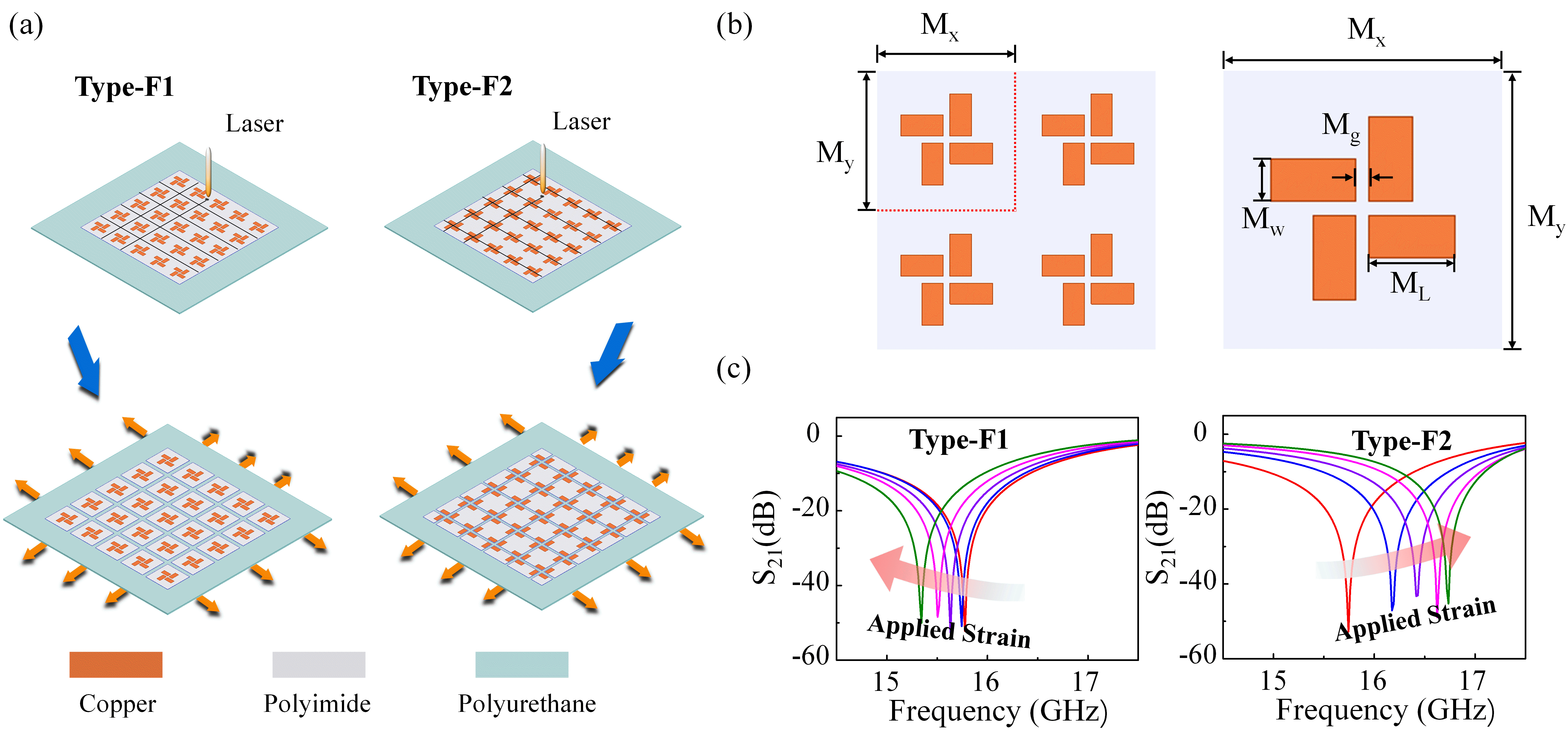}
	\caption{(a) Schematic of block definition design process to realize the stretchable MMs with configurable resonant frequency shifts. (b) Geometric illustration of the precursor structure used in the block definition of type-F1 and type-F2 MMs. (c) Simulation results of transmission coefficients of type-F1 and type-F2 MMs with different applied strains.}
	\label{figure6}
\end{figure}
\begin{table}[!h]
	%% increase table row spacing, adjust to taste
	\renewcommand{\arraystretch}{1.1}
	% if using array.sty, it might be a good idea to tweak the value of
	\caption{Parameters of the MM consisted of rectangle blocks}
	\label{table_fss2}
	\centering
	\begin{tabular}{|c|c|c|c|c|}
		\hline
		$M_x$  & $M_y$  & $M_g$ & $M_w$ &$M_L$  \\
		\hline
		15.0 $mm$& 15.0 $mm$& 0.1 $mm$ & 3.0 $mm$ & 6.0 $mm$  \\
		\hline
	\end{tabular}
\end{table}

The MM is partitioned into two types (type-F1 and type-F2) with identical precursor structures using ultraviolet laser micromachining, as depicted in Figure 6(a). The precursor structure of the MM consists of periodic unit cells created by four rectangle strips with C-4 rotation symmetry, as shown in Figure 6(b), and the structural parameters are given in Table \ref{table_fss2}. Figure 6(c) shows the simulation results of the transmission spectra of both types of MMs under biaxial tensile strain. The two types of MMs exhibit opposite responses to deformation with the same applied strain. For type-F1 MM, the resonant frequency shows a red shift from 15.78 GHz to 15.34 GHz, while the resonant frequency of type-F2 MM increases from 15.78 GHz to 16.74 GHz with the same tensile strain. The observed opposite relation between the deformation and resonant frequency can be attributed to the coupling condition determined by the block arrangement. The block definition used in type-F1 MM partitions the precursor structure to allow the relative position of the four strips to remain stable with the applied strain. The deformation of type-F1 MM only leads to a decrease in coupling strength between the blocks, which is much weaker than the coupling between the strips. In contrast, the cutting silts used to define the type-F2 MM are located in the middle of the four strips in each unit cell, where the coupling between the strips is strong. When tensile strain is applied, the coupling capacitance of the type-F2 MM is significantly reduced, causing the resonant frequency to exhibit a blue shift.

The utilization of the EIT phenomenon has been proven effective in achieving slow light devices and enhanced non-linear effects, as demonstrated in various studies \cite{wang2020electromagnetically,hu2020ultrafast,jung2019electrical,wang2019electromagnetically,nakanishi2018storage,ru2020talbot,guo2020transient}. The bandwidth of the transmission window is a crucial parameter for the MMs used in EIT-based applications \cite{yahiaoui2017active}, and the block definition design can be used to obtain MMs with configurable dependence of transmission bandwidth on deformation, using the same precursor structure. 
\begin{figure}[!h]
	\centering
	\includegraphics[width=4.5in,height=3.2in]{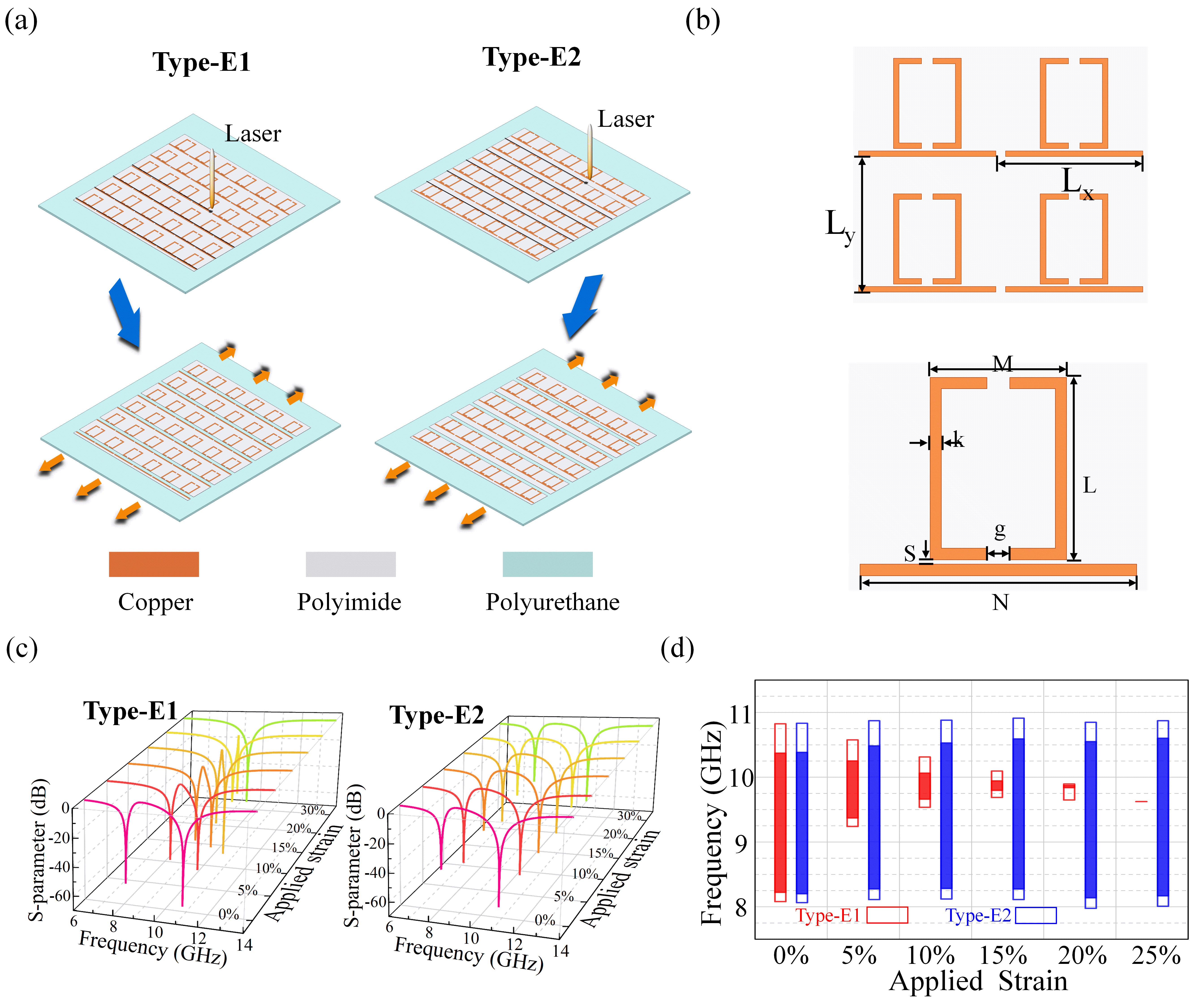}
	\caption{(a) Schematic of block definition design process to realize the stretchable MMs with configurable EIT effect. (b) Geometric illustration of the MM. (c) Transmission coefficients of the type-E1 and type-E2 MMs with different applied strains. (d) Simulation results of the dependence of resonant frequency and bandwidth on the applied strain. The internal color bars represent cut-off frequencies at -10 dB, and the top and bottom lines represent the two resonant dips of the EIT effect.}
	\label{figure7}
\end{figure}

In this regard, two types of MMs, type-E1 and type-E2, were defined with a precursor structure consisting of periodically arranged cut-wire and split-ring elements, as shown in Figure 7(a). The structural details of the MMs are illustrated in Figure 7(b), and the corresponding parameters are listed in Table \ref{table_EIT}. The simulation results of the transmission spectra of the two types of MMs with different applied tensile strains are shown in Figure 7(c). It is evident from the transmission spectra that there is a significant difference in the sensitivity of the transmission bandwidth to deformation between type-E1 and type-E2 MMs. The transmission bandwidth of type-E1 MM demonstrates a very sensitive response to the applied strain, indicating its potential as a mechanical-tunable MM for EIT applications. On the other hand, the transmission bandwidth of type-E2 MM remains almost the same during deformation, making it a good candidate for conformal integration. The dependence of the resonant frequencies and bandwidth of the MMs are illustrated in detail in Figure 7(d). It is observed that both type-E1 and type-E2 MMs have similar transmission characteristics with a bandwidth of 2.7 GHz at the initial state (applied strain = 0\%). However, the transmission bandwidth of type-E1 MM gradually decreases from 2.7 GHz with increasing applied strain, while the transmission bandwidth of type-E2 MM remains constant with applied strain ranging from 0\% to 25\%. These results demonstrate the potential of the block definition design in configuring the dependence of MM's properties on deformation, thereby indicating its applicability to MM for various applications.
\begin{table}[!h]
	%% increase table row spacing, adjust to taste
	\renewcommand{\arraystretch}{1.1}
	% if using array.sty, it might be a good idea to tweak the value of
	\caption{Parameters of the EIT-MM}
	\label{table_EIT}
	\centering
	\begin{tabular}{|c|c|c|c|c|c|c|c|}
		\hline
		$L_x$  & $L_y$  & $M$ & $N$ & $L$  & $g$ & $s$  & $k$  \\\hline
        5.0 $mm$& 5.0 $mm$&6.0 $mm$ &12.1 $mm$ & 8.0 $mm$ &	1.0	$mm$& 0.2 $mm$& 0.5 $mm$  \\\hline
		
	\end{tabular}
\end{table}

To clarify the advantages of our proposed approach, we conducted a comprehensive comparison with previously reported methods for fabricating stretchable MMs. The comparison results are presented in Table \ref{table_comparasion}. It is noteworthy that our transfer-define design for MMs can achieve both conformal integration and mechanical reconfiguration using only one type of precursor structure, while other methods can only achieve one of these functions with one precursor structure. Moreover, our proposed method enables the production of large-scale MMs (up to $m^2$) through compatible and convenient fabrication processes, which is crucial for their application in the radio-frequency regime.
\begin{table*}[!ht]\scriptsize
	\renewcommand\arraystretch{1.2}
	\caption{Comparison between the proposed method and previous studies.}
	\label{table_comparasion}
	\centering
	\setlength{\leftskip}{-30pt}
	\setlength{\tabcolsep}{1.2mm}{
		\begin{tabular}{clllllp{2.5cm}}
			\toprule
			Ref. & Methods & Frequency  & Size  & Flexibility & Function & Applications \\
			\midrule
			\cite{khodasevych2012elastomeric} & Photolithography & THz & Limited(cm$^2$) & Bendable & High-quality magnetic resonance & Conformal integration \\
			\cite{choi2011terahertz} & Photolithography & THz & Limited(cm$^2$) & Bendable & High refractive index & Conformal integration \\
			\cite{li2013mechanically}  &Photolithography	& THz  & Limited(cm$^2$) & Stretchable & Frequency tuning & Frequency reconfiguring  \\ 	     
			\cite{chen2019flexible} &Printed circuit board technology & GHz  &  Large(m$^2$) & Bendable & Electromagnetic shielding & Conformal integration \\             
			\cite{bodaghi2017large} &3D direct writing  & GHz  & Large(dm$^2$) & Stretchable & Frequency tuning  &  Frequency reconfiguring \\			 
			\cite{phon2021mechanical} &3D direct writing  & GHz  & Large(dm$^2$) & Stretchable & Beam splitting and steering & Mode reconfiguring \\		   	  
			\cite{zhou2018stretchable}&	Screen printing  & GHz &  Large(m$^2$)  &  Stretchable  & Microwave absorber & Frequency reconfiguring\\
			\cite{yoo2020stretchable} &	Screen printing	& GHz & Large(m$^2$)  &  Stretchable & Strain sensing & Frequency reconfiguring \\
			\cite{yang2016flexible} &	Micro-channel   & GHz   &  Large(m$^2$) & Stretchable & Electromagnetic shielding & Frequency reconfiguring\\
			\cite{pryce2010highly} &	Electron beam lithography & Optical  & Limited(mm$^2$)  & Stretchable & Large frequency tunability & Frequency reconfiguring \\
			\cite{gholipour2017organometallic}& 	Focused ion beam   & Optical  &  Limited(mm$^2$) &  Bendable &  Structural color & Conformal integration \\
			\cite{chanda2011large} &	Nanoimprint  & Optical  &  Limited(mm$^2$)  &  Bendable & Negative index  & Conformal integration  \\
			\cite{lee2016heterogeneously} & Modular transfer   & THz   &  Limited(cm$^2$)  &  Bendable & Frequency selecting  & Conformal integration \\
			\cite{lee2012reversibly} &	Surface-relief assisted transfer & THz  &  Limited(cm$^2$) &  Stretchable  & Frequency tuning  & Frequency reconfiguring\\
			\cite{fan2020mechanical} &	Surface-relief assisted transfer & THz  &  Limited(cm$^2$) &  Stretchable  & Frequency tuning & Frequency reconfiguring \\
			\cite{ee2016tunable} &	Kinetically controlled transfer & Optical  &  Limited(mm$^2$)  &  Stretchable & Optical zoom lens & Phase reconfiguring\\
			\cite{malek2017strain} &	Kinetically controlled transfer & Optical  &  Limited(mm$^2$)  &  Stretchable & Holograms & Phase reconfiguring \\
			\cite{zhang2021reconfigurable} &Kinetically controlled transfer & Optical  &  Limited(mm$^2$)  &  Stretchable &  Image processing size & Phase reconfiguring \; \; \; \; Frequency reconfiguring \\            	 	
			This work&	Transfer-define   &  GHz   &  Large(m$^2$) &  Stretchable & Frequency and bandwidth tuning & Conformal integration \; \; \; \; Frequency reconfiguring\\
			\bottomrule
		\end{tabular}
	}
\end{table*}

\section{Conclusion}
In this work, a block definition design for the stretchable MMs is proposed. Instead of directly transferring the MMs to soft substrate, the introduction of “transfer-define” fabrication processes in the block definition design allows the formation of MMs by cutting precursor laminate into periodical blocks after transferring the precursor in a whole to the soft substrate. The “transfer-define” fabrication process avoids the transfer of deformable structure and eliminates the risk of misalignment, which enables the realization of large-scale MMs for radio-frequency applications. The MMs with block definitions of one column, two columns and three columns of square-loop elements in each block are firstly designed and fabricated with the block definition design respectively. By evaluating the transmission spectra of the MMs by simulation and experimental measurement, it's demonstrated that the sensitivity of the MMs to the deformation can be effectively configured with the implementation of different block definitions. Then, we also designed two other demos with the block definition design for exhibiting its wide application prospect, another stretchable MM with configurable resonant frequency shift and an MM with configurable EIT effect. By defining the precursor structure into two different block arrangement, the MM design show positive/negative correlation between resonant frequency and tensile deformation, and the EIT MM can configure the transmission window to be sensitive/insensitive to the applied tensile strain. Thus, the proposed block definition design for the stretchable MMs is expected to provide a highly flexible and convenient way to realize the stretchable MM with different response to deformation, and promote the application of stretchable MM in various scenarios from conformal integration to mechanical modulation.

\begin{backmatter}
\bmsection{Funding}
National Natural Science Foundation of China (Nos. 61825102, 52021001, and 61901085); Medico-Engineering Cooperation Funds from University of Electronic Science and Technology of China (No. ZYGX2021YGLH008)
	
\bmsection{Disclosures}
The authors declare that there are no conflicts of interest related to this article.
	
\bmsection{Data Availability}
Data underlying the results presented in this paper are available from the corresponding authors
upon reasonable request.
\end{backmatter}
\bibliography{references}

\begin{thebibliography}{10}
\newcommand{\enquote}[1]{``#1''}

\bibitem{landy2008perfect}
N.~I. Landy, S.~Sajuyigbe, J.~J. Mock, D.~R. Smith, and W.~J. Padilla,
  \enquote{Perfect metamaterial absorber,} {\protect\JournalTitle{Physical
  Review Letters}} \textbf{100}, 207402 (2008).

\bibitem{liu2010infrared}
N.~Liu, M.~Mesch, T.~Weiss, M.~Hentschel, and H.~Giessen, \enquote{Infrared
  perfect absorber and its application as plasmonic sensor,}
  {\protect\JournalTitle{Nano Letters}} \textbf{10}, 2342--2348 (2010).

\bibitem{tittl2015switchable}
A.~Tittl, A.-K.~U. Michel, M.~Sch{\"a}ferling, X.~Yin, B.~Gholipour, L.~Cui,
  M.~Wuttig, T.~Taubner, F.~Neubrech, and H.~Giessen, \enquote{A switchable
  mid-infrared plasmonic perfect absorber with multispectral thermal imaging
  capability,} {\protect\JournalTitle{Advanced Materials}} \textbf{27},
  4597--4603 (2015).

\bibitem{durmaz2019polarization}
H.~Durmaz, A.~E. Cetin, Y.~Li, and R.~Paiella, \enquote{A polarization
  insensitive wide-band perfect absorber,} {\protect\JournalTitle{Advanced
  Engineering Materials}} \textbf{21}, 1900188 (2019).

\bibitem{wang2019large}
S.~Wang, F.~Chen, R.~Ji, M.~Hou, F.~Yi, W.~Zheng, T.~Zhang, and W.~Lu,
  \enquote{Large-area low-cost dielectric perfect absorber by one-step
  sputtering,} {\protect\JournalTitle{Advanced Optical Materials}} \textbf{7},
  1801596 (2019).

\bibitem{schurig2006metamaterial}
D.~Schurig, J.~J. Mock, B.~Justice, S.~A. Cummer, J.~B. Pendry, A.~F. Starr,
  and D.~R. Smith, \enquote{Metamaterial electromagnetic cloak at microwave
  frequencies,} {\protect\JournalTitle{Science}} \textbf{314}, 977--980 (2006).

\bibitem{zhang2019phase}
F.~Zhang, C.~Li, Y.~Fan, R.~Yang, N.-H. Shen, Q.~Fu, W.~Zhang, Q.~Zhao,
  J.~Zhou, T.~Koschny \emph{et~al.}, \enquote{Phase-modulated scattering
  manipulation for exterior cloaking in metal--dielectric hybrid
  metamaterials,} {\protect\JournalTitle{Advanced Materials}} \textbf{31},
  1903206 (2019).

\bibitem{qian2020deep}
C.~Qian, B.~Zheng, Y.~Shen, L.~Jing, E.~Li, L.~Shen, and H.~Chen,
  \enquote{Deep-learning-enabled self-adaptive microwave cloak without human
  intervention,} {\protect\JournalTitle{Nature Photonics}} \textbf{14},
  383--390 (2020).

\bibitem{wan2020holographic}
W.~Wan, W.~Qiao, D.~Pu, R.~Li, C.~Wang, Y.~Hu, H.~Duan, L.~J. Guo, and L.~Chen,
  \enquote{Holographic sampling display based on metagratings,}
  {\protect\JournalTitle{Iscience}} \textbf{23}, 100773 (2020).

\bibitem{cui2020information}
T.~J. Cui, L.~Li, S.~Liu, Q.~Ma, L.~Zhang, X.~Wan, W.~X. Jiang, and Q.~Cheng,
  \enquote{Information metamaterial systems,} {\protect\JournalTitle{Iscience}}
  \textbf{23} (2020).

\bibitem{guo2019reconfigurable}
J.~Guo, T.~Wang, H.~Zhao, X.~Wang, S.~Feng, P.~Han, W.~Sun, J.~Ye, G.~Situ,
  H.-T. Chen \emph{et~al.}, \enquote{Reconfigurable terahertz metasurface pure
  phase holograms,} {\protect\JournalTitle{Advanced Optical Materials}}
  \textbf{7}, 1801696 (2019).

\bibitem{jiang2011conformal}
Z.~H. Jiang, S.~Yun, F.~Toor, D.~H. Werner, and T.~S. Mayer, \enquote{Conformal
  dual-band near-perfectly absorbing mid-infrared metamaterial coating,}
  {\protect\JournalTitle{ACS Nano}} \textbf{5}, 4641--4647 (2011).

\bibitem{cong2014highly}
L.~Cong, N.~Xu, J.~Gu, R.~Singh, J.~Han, and W.~Zhang, \enquote{Highly flexible
  broadband terahertz metamaterial quarter-wave plate,}
  {\protect\JournalTitle{Laser \& Photonics Reviews}} \textbf{8}, 626--632
  (2014).

\bibitem{zhou2018stretchable}
P.~Zhou, L.~Wang, G.~Zhang, J.~Jiang, H.~Chen, Y.~Zhou, D.~Liang, and L.~Deng,
  \enquote{A stretchable metamaterial absorber with deformation compensation
  design at microwave frequencies,} {\protect\JournalTitle{IEEE Transactions on
  Antennas and Propagation}} \textbf{67}, 291--297 (2018).

\bibitem{hashemi2019flexible}
M.~R.~M. Hashemi, A.~C. Fikes, M.~Gal-Katziri, B.~Abiri, F.~Bohn,
  A.~Safaripour, M.~D. Kelzenberg, E.~L. Warmann, P.~Espinet, N.~Vaidya
  \emph{et~al.}, \enquote{A flexible phased array system with low areal mass
  density,} {\protect\JournalTitle{Nature Electronics}} \textbf{2}, 195--205
  (2019).

\bibitem{lee2012reversibly}
S.~Lee, S.~Kim, T.-T. Kim, Y.~Kim, M.~Choi, S.~H. Lee, J.-Y. Kim, and B.~Min,
  \enquote{Reversibly stretchable and tunable terahertz metamaterials with
  wrinkled layouts,} {\protect\JournalTitle{Advanced Materials}} \textbf{24},
  3491--3497 (2012).

\bibitem{liang2015anomalous}
L.~Liang, M.~Qi, J.~Yang, X.~Shen, J.~Zhai, W.~Xu, B.~Jin, W.~Liu, Y.~Feng,
  C.~Zhang \emph{et~al.}, \enquote{Anomalous terahertz reflection and
  scattering by flexible and conformal coding metamaterials,}
  {\protect\JournalTitle{Advanced Optical Materials}} \textbf{3}, 1374--1380
  (2015).

\bibitem{yang2016flexible}
S.~Yang, P.~Liu, M.~Yang, Q.~Wang, J.~Song, and L.~Dong, \enquote{From flexible
  and stretchable meta-atom to metamaterial: A wearable microwave meta-skin
  with tunable frequency selective and cloaking effects,}
  {\protect\JournalTitle{Scientific Reports}} \textbf{6}, 21921 (2016).

\bibitem{ee2016tunable}
H.-S. Ee and R.~Agarwal, \enquote{Tunable metasurface and flat optical zoom
  lens on a stretchable substrate,} {\protect\JournalTitle{Nano Letters}}
  \textbf{16}, 2818--2823 (2016).

\bibitem{Gurrala2017Fully}
P.~Gurrala, S.~Oren, P.~Liu, J.~Song, and L.~Dong, \enquote{Fully conformal
  square-patch frequency-selective surface toward wearable electromagnetic
  shielding,} {\protect\JournalTitle{IEEE Antennas and Wireless Propagation
  Letters}} \textbf{16}, 2602--2605 (2017).

\bibitem{malek2017strain}
S.~C. Malek, H.-S. Ee, and R.~Agarwal, \enquote{Strain multiplexed metasurface
  holograms on a stretchable substrate,} {\protect\JournalTitle{Nano Letters}}
  \textbf{17}, 3641--3645 (2017).

\bibitem{nauroze2018continuous}
S.~A. Nauroze, L.~S. Novelino, M.~M. Tentzeris, and G.~H. Paulino,
  \enquote{Continuous-range tunable multilayer frequency-selective surfaces
  using origami and inkjet printing,} {\protect\JournalTitle{Proceedings of the
  National Academy of Sciences}} \textbf{115}, 13210--13215 (2018).

\bibitem{pryce2010highly}
I.~M. Pryce, K.~Aydin, Y.~A. Kelaita, R.~M. Briggs, and H.~A. Atwater,
  \enquote{Highly strained compliant optical metamaterials with large frequency
  tunability,} {\protect\JournalTitle{Nano Letters}} \textbf{10}, 4222--4227
  (2010).

\bibitem{lee2019single}
S.~Lee, W.~T. Kim, J.-H. Kang, B.~J. Kang, F.~Rotermund, and Q.-H. Park,
  \enquote{Single-layer metasurfaces as spectrally tunable terahertz half-and
  quarter-waveplates,} {\protect\JournalTitle{ACS Applied Materials \&
  Interfaces}} \textbf{11}, 7655--7660 (2019).

\bibitem{choi2016electroactive}
J.-H. Choi, J.~Ahn, J.-B. Kim, Y.-C. Kim, J.-Y. Lee, and I.-K. Oh, \enquote{An
  electroactive, tunable, and frequency selective surface utilizing highly
  stretchable dielectric elastomer actuators based on functionally antagonistic
  aperture control,} {\protect\JournalTitle{Small}} \textbf{12}, 1840--1846
  (2016).

\bibitem{kamali2016decoupling}
S.~M. Kamali, A.~Arbabi, E.~Arbabi, Y.~Horie, and A.~Faraon,
  \enquote{Decoupling optical function and geometrical form using conformal
  flexible dielectric metasurfaces,} {\protect\JournalTitle{Nature
  Communications}} \textbf{7}, 1--7 (2016).

\bibitem{yu2013stretchable}
C.~L. Yu, H.~Kim, N.~De~Leon, I.~W. Frank, J.~T. Robinson, M.~McCutcheon,
  M.~Liu, M.~D. Lukin, M.~Loncar, and H.~Park, \enquote{Stretchable photonic
  crystal cavity with wide frequency tunability,} {\protect\JournalTitle{Nano
  Letters}} \textbf{13}, 248--252 (2013).

\bibitem{ni2015ultrathin}
X.~Ni, Z.~J. Wong, M.~Mrejen, Y.~Wang, and X.~Zhang, \enquote{An ultrathin
  invisibility skin cloak for visible light,} {\protect\JournalTitle{Science}}
  \textbf{349}, 1310--1314 (2015).

\bibitem{geiger2020flexible}
S.~Geiger, J.~Michon, S.~Liu, J.~Qin, J.~Ni, J.~Hu, T.~Gu, and N.~Lu,
  \enquote{Flexible and stretchable photonics: the next stretch of
  opportunities,} {\protect\JournalTitle{ACS Photonics}}  (2020).

\bibitem{yilmaz2009design}
A.~E. Yilmaz and M.~Kuzuoglu, \enquote{Design of the square loop frequency
  selective surfaces with particle swarm optimization via the equivalent
  circuit model.} {\protect\JournalTitle{Radioengineering}} \textbf{18} (2009).

\bibitem{parker1991gentleman}
E.~A. Parker, \enquote{The gentleman's guide to frequency selective surfaces,}
  (1991).

\bibitem{monacelli2005infrared}
B.~Monacelli, J.~B. Pryor, B.~A. Munk, D.~Kotter, and G.~D. Boreman,
  \enquote{Infrared frequency selective surface based on circuit-analog square
  loop design,} {\protect\JournalTitle{IEEE Transactions on Antennas and
  Propagation}} \textbf{53}, 745--752 (2005).

\bibitem{wang2019strain}
F.~Wang, S.~Chen, Q.~Wu, R.~Zhang, and P.~Sun, \enquote{Strain-induced
  structural and dynamic changes in segmented polyurethane elastomers,}
  {\protect\JournalTitle{Polymer}} \textbf{163}, 154--161 (2019).

\bibitem{dal2021performance}
H.~Dal, K.~A{\c{c}}{\i}kg{\"o}z, and Y.~Badienia, \enquote{On the performance
  of isotropic hyperelastic constitutive models for rubber-like materials: a
  state of the art review,} {\protect\JournalTitle{Applied Mechanics Reviews}}
  \textbf{73} (2021).

\bibitem{wang2020electromagnetically}
C.~Wang, X.~Jiang, G.~Zhao, M.~Zhang, C.~W. Hsu, B.~Peng, A.~D. Stone,
  L.~Jiang, and L.~Yang, \enquote{Electromagnetically induced transparency at a
  chiral exceptional point,} {\protect\JournalTitle{Nature Physics}}
  \textbf{16}, 334--340 (2020).

\bibitem{hu2020ultrafast}
Y.~Hu, T.~Jiang, H.~Sun, M.~Tong, J.~You, X.~Zheng, Z.~Xu, and X.~Cheng,
  \enquote{Ultrafast frequency shift of electromagnetically induced
  transparency in terahertz metaphotonic devices,} {\protect\JournalTitle{Laser
  \& Photonics Reviews}} \textbf{14}, 1900338 (2020).

\bibitem{jung2019electrical}
H.~Jung, H.~Jo, W.~Lee, B.~Kim, H.~Choi, M.~S. Kang, and H.~Lee,
  \enquote{Electrical control of electromagnetically induced transparency by
  terahertz metamaterial funneling,} {\protect\JournalTitle{Advanced Optical
  Materials}} \textbf{7}, 1801205 (2019).

\bibitem{wang2019electromagnetically}
Q.~Wang, L.~Yu, H.~Gao, S.~Chu, and W.~Peng, \enquote{Electromagnetically
  induced transparency in an all-dielectric nano-metamaterial for slow light
  application,} {\protect\JournalTitle{Optics Express}} \textbf{27},
  35012--35026 (2019).

\bibitem{nakanishi2018storage}
T.~Nakanishi and M.~Kitano, \enquote{Storage and retrieval of electromagnetic
  waves using electromagnetically induced transparency in a nonlinear
  metamaterial,} {\protect\JournalTitle{Applied Physics Letters}} \textbf{112},
  201905 (2018).

\bibitem{ru2020talbot}
J.-M. Ru, Z.-K. Wu, Y.-G. Zhang, F.~Wen, and Y.-Z. Gu, \enquote{Talbot effect
  in nonparaxial self-accelerating beams with electromagnetically induced
  transparency,} {\protect\JournalTitle{Frontiers of Physics}} \textbf{15},
  1--7 (2020).

\bibitem{guo2020transient}
Y.-W. Guo, S.-L. Xu, J.-R. He, P.~Deng, M.~R. Beli{\'c}, and Y.~Zhao,
  \enquote{Transient optical response of cold rydberg atoms with
  electromagnetically induced transparency,} {\protect\JournalTitle{Physical
  Review A}} \textbf{101}, 023806 (2020).

\bibitem{yahiaoui2017active}
R.~Yahiaoui, M.~Manjappa, Y.~K. Srivastava, and R.~Singh, \enquote{Active
  control and switching of broadband electromagnetically induced transparency
  in symmetric metadevices,} {\protect\JournalTitle{Applied Physics Letters}}
  \textbf{111}, 021101 (2017).

\bibitem{khodasevych2012elastomeric}
I.~Khodasevych, C.~M. Shah, S.~Sriram, M.~Bhaskaran, W.~Withayachumnankul,
  B.~Ung, H.~Lin, W.~Rowe, D.~Abbott, and A.~Mitchell, \enquote{Elastomeric
  silicone substrates for terahertz fishnet metamaterials,}
  {\protect\JournalTitle{Applied Physics Letters}} \textbf{100}, 061101 (2012).

\bibitem{choi2011terahertz}
M.~Choi, S.~H. Lee, Y.~Kim, S.~B. Kang, J.~Shin, M.~H. Kwak, K.-Y. Kang, Y.-H.
  Lee, N.~Park, and B.~Min, \enquote{A terahertz metamaterial with unnaturally
  high refractive index,} {\protect\JournalTitle{Nature}} \textbf{470},
  369--373 (2011).

\bibitem{li2013mechanically}
J.~Li, C.~M. Shah, W.~Withayachumnankul, B.~S.-Y. Ung, A.~Mitchell, S.~Sriram,
  M.~Bhaskaran, S.~Chang, and D.~Abbott, \enquote{Mechanically tunable
  terahertz metamaterials,} {\protect\JournalTitle{Applied Physics Letters}}
  \textbf{102}, 121101 (2013).

\bibitem{chen2019flexible}
S.~Chen, T.~Pan, Z.~Yan, L.~Dai, Y.~Peng, M.~Gao, and Y.~Lin, \enquote{Flexible
  serpentinelike frequency selective surface for conformal applications with
  stable frequency response,} {\protect\JournalTitle{IEEE Antennas and Wireless
  Propagation Letters}} \textbf{18}, 1477--1481 (2019).

\bibitem{bodaghi2017large}
M.~Bodaghi, A.~Damanpack, G.~Hu, and W.~Liao, \enquote{Large deformations of
  soft metamaterials fabricated by 3d printing,}
  {\protect\JournalTitle{Materials \& Design}} \textbf{131}, 81--91 (2017).

\bibitem{phon2021mechanical}
R.~Phon, Y.~Kim, E.~Park, H.~Jeong, and S.~Lim, \enquote{Mechanical and
  self-deformable spatial modulation beam steering and splitting metasurface,}
  {\protect\JournalTitle{Advanced Optical Materials}} \textbf{9}, 2100821
  (2021).

\bibitem{yoo2020stretchable}
Y.~Yoo, H.~Jeong, D.~Lim, and S.~Lim, \enquote{Stretchable screen-printed
  metasurfaces for wireless strain sensing applications,}
  {\protect\JournalTitle{Extreme Mechanics Letters}} \textbf{41}, 100998
  (2020).

\bibitem{gholipour2017organometallic}
B.~Gholipour, G.~Adamo, D.~Cortecchia, H.~N. Krishnamoorthy, M.~D. Birowosuto,
  N.~I. Zheludev, and C.~Soci, \enquote{Organometallic perovskite
  metasurfaces,} {\protect\JournalTitle{Advanced Materials}} \textbf{29},
  1604268 (2017).

\bibitem{chanda2011large}
D.~Chanda, K.~Shigeta, S.~Gupta, T.~Cain, A.~Carlson, A.~Mihi, A.~J. Baca,
  G.~R. Bogart, P.~Braun, and J.~A. Rogers, \enquote{Large-area flexible 3d
  optical negative index metamaterial formed by nanotransfer printing,}
  {\protect\JournalTitle{Nature Nanotechnology}} \textbf{6}, 402--407 (2011).

\bibitem{lee2016heterogeneously}
S.~Lee, B.~Kang, H.~Keum, N.~Ahmed, J.~A. Rogers, P.~M. Ferreira, S.~Kim, and
  B.~Min, \enquote{Heterogeneously assembled metamaterials and metadevices via
  3d modular transfer printing,} {\protect\JournalTitle{Scientific Reports}}
  \textbf{6}, 1--11 (2016).

\bibitem{fan2020mechanical}
X.~Fan, Y.~Li, S.~Chen, Y.~Xing, and T.~Pan, \enquote{Mechanical terahertz
  modulation by skin-like ultrathin stretchable metasurface,}
  {\protect\JournalTitle{Small}} \textbf{16}, 2002484 (2020).

\bibitem{zhang2021reconfigurable}
X.~Zhang, Y.~Zhou, H.~Zheng, A.~E. Linares, F.~C. Ugwu, D.~Li, H.-B. Sun,
  B.~Bai, and J.~G. Valentine, \enquote{Reconfigurable metasurface for image
  processing,} {\protect\JournalTitle{Nano Letters}} \textbf{21}, 8715--8722
  (2021).

\end{thebibliography}
\end{document}